\documentclass[aps,pra,twocolumn,tightenlines,floatfix,showpacs,superscriptaddress,10pt]{revtex4-1}
\usepackage{upgreek}
\usepackage{epsfig}
\usepackage{tikz}
\usepackage{graphicx}
\usepackage{epstopdf}
\usepackage{fancybox}
\usepackage{makeidx}
\usepackage{mathrsfs}
\usepackage{appendix}
\usepackage{amsmath}
\usepackage{amsfonts}
\usepackage{natbib}
\usepackage[figuresright]{rotating}
\usepackage{floatpag}
\usepackage{natbib}
\usepackage{marginnote}
\usepackage{enumerate}
\usepackage{mathtools}

\newcommand{\bra}[1]{\langle #1 |}
\newcommand{\ket}[1]{| #1 \rangle}

\begin{document}

\title{Digital Memcomputing: from Logic to Dynamics to Topology}

\author{Massimiliano Di Ventra}
\email{diventra@physics.ucsd.edu} 
\affiliation{Department of Physics, University of California San Diego, La Jolla, California 92093, USA}
\author{Igor V. Ovchinnikov} 
\email{email: igor.vlad.ovchinnikov@gmail.com}
\affiliation{Electrical Engineering Department, University of California, Los Angeles, 90095 CA}

\begin{abstract} 
Digital memcomputing machines (DMMs) are a class of computational machines designed to solve combinatorial optimization problems. A practical realization of DMMs can be accomplished via electrical circuits of highly non-linear, point-dissipative dynamical systems engineered so that periodic orbits and chaos can be avoided. A given logic problem is first mapped into this type of dynamical system whose point attractors represent the solutions of the original problem. A DMM then finds the solution via a succession of elementary instantons whose role is to eliminate solitonic configurations of logical inconsistency (``logical defects'') from the circuit. By employing a supersymmetric theory of dynamics, a DMM can be described by a cohomological field theory that allows for computation of certain topological matrix elements on instantons that have the mathematical meaning of intersection numbers on instantons. We discuss the ``dynamical'' meaning of these matrix elements, and argue that the number of elementary instantons needed to reach the solution cannot exceed the number of state variables of DMMs, which in turn can only grow at most polynomially with the size of the problem. These results shed further light on the relation between logic, dynamics and topology in digital memcomputing. 
\end{abstract}

\maketitle

\section{Introduction}\label{intro}

{\it Memcomputing}~\cite{diventra13a} is a novel computing paradigm in which computation is done by the memory and in memory~\cite{UMM}. A realization of this concept is represented by {\it digital} memcomputing machines (DMMs) as recently proposed in~\cite{DMM2,DMMperspective}, where a given combinatorial optimization problem is first mapped into a dynamical system whose point attractors represent the solutions of the original problem. Despite operating in continuous time, DMMs map a finite string of symbols (e.g., 0s and 1s) into a finite string of symbols, thus representing scalable machines~\cite{DMM2,DMMperspective}.

\begin{figure}[t!]
	\includegraphics[width=1.0\columnwidth]{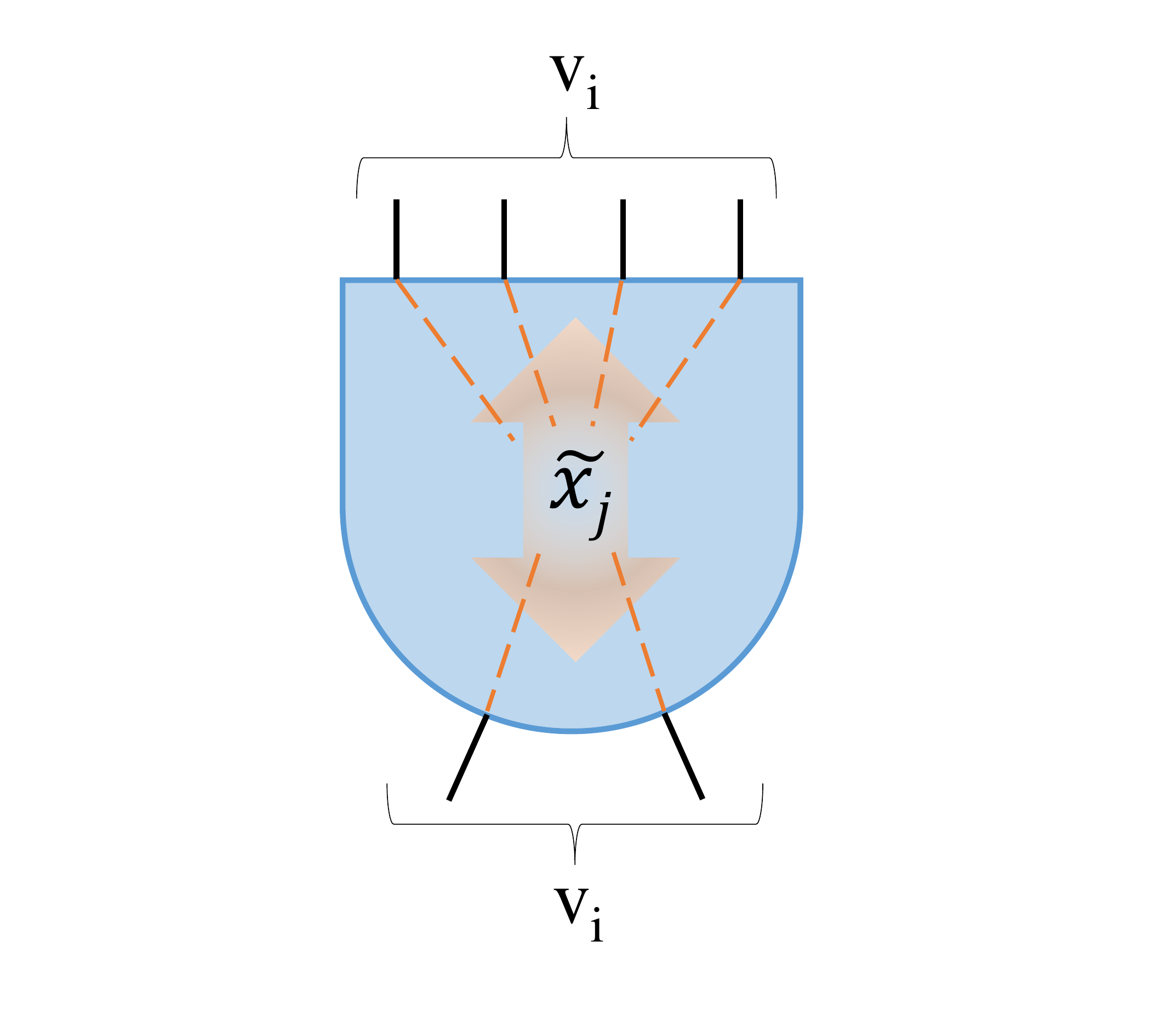}
	\caption{Schematic of a self-organizing logic gate with internal memory variables, ${\tilde x}_j$, that mediate the interaction among 
		the external voltage variables $v_i$.} 
	\label{fig1}
\end{figure}

Although not unique, a practical realization of a DMM can be realized with an electrical circuit that has the same skeleton as the underlying logical problem it needs to solve. The logic gates of this circuit, called self-organizing logic gates (SOLGs)~\footnote{Or self-organizing algebraic gates, if algebraic relations need to be satisfied~\cite{ILP}}, are such that they are at rest only if the voltages on their terminals are logically consistent with the logic operation each gate is responsible for (see Fig.~\ref{fig1})~\cite{DMM2,DMMperspective}.

The catch so far is that a collection of SOLGs, viewed as a dynamical system, may then have additional attractors that do not correspond to solutions of the 
combinatorial optimization problem at hand. This is where time non-locality, represented by internal slow ``memory'' variables, comes into play~\footnote{From an electrical engineering point of view, DMM circuits have both passive (such as resistors, capacitors, etc.) and active (such as transistors) components. The internal degrees of freedom are slow providing time non-locality to the system and they are practically introduced by means of circuit elements with memory (such as memristive or memcapacitive elements)~\cite{09_memelements}, or emulated by a combination of active elements~\cite{Pershin2010}.}. 

From a logical point of view, the role of these variables is to relax the ``digitalization'' constraints on the voltages of each terminal of the SOLGs, and allow them to go through an ``analog'' transient regime during the solution search when they are not constrained to be integers. From a dynamical-system point of view, the presence of these memory variables extends the dimensionality of the phase space as compared to that spanned by the logical voltages only. It is in this higher-dimensional phase space that it is easier for the dynamical system to find its way to the solution by avoiding being trapped in any spurious state~\cite{DMM2,DMMperspective}. A nontrivial part of the theory of DMMs is precisely to prove that solutions of the given logical problem are the only stable minima of the corresponding electrical circuit. The details of this proof can be found in Ref.~\cite{DMM2}, and will not 
be discussed further in this paper whose role is to explore the relation between logic, dynamics and topology.

The equations of motion for DMMs are, of course, specific to the given combinatorial optimization problem \footnote{See, e.g., Ref.~\cite{DMM2} for explicit equations that solve the factorization and the subset-sum problem or Ref.~\cite{spinglass} for the Ising spin glass}. For the purpose of this paper, however, it suffices to view these equations of motion from the most general perspective, i.e., as non-linear (autonomous) ordinary differential equations of the type:
\begin{equation}
\dot x(t) = F(x(t)) \label{ODE},
\end{equation}
where $x$ is a $D$-dimensional set of variables that defines the state of the system, and belongs to some $D$-dimensional topological manifold, $X \subset \mathbb{R}^D$, called phase space, and $F$ is the flow vector field that represents the dynamics of $x$ according to the circuit that solves the problem at hand. In the following, we will assume that once the combinatorial optimization problem to solve has been given and transformed into a physical problem described by a set of Eqs.~(\ref{ODE}), the vector $x$ contains, say, $m$ voltages, $v_i$, at all terminals of the corresponding electrical circuit, and $n$ internal memory variables $\tilde x_j$ that provide memory to the system (with $D=n+m$).  

In fact, Eqs.~(\ref{ODE}) must have an additional mathematical property to represent a valid DMM: they must describe {\it point-dissipative systems}~\cite{hale_2010_asymptotic}. These are special dynamical systems that support a compact global
attractor, so that all trajectories of the system are {\it bounded} and will eventually end up into the global attractor, irrespective of the initial conditions. 
A dynamical system~(\ref{ODE}) with this property can then be designed to avoid chaos~\cite{no-chaos} and periodic orbits~\cite{noperiod}. This means that 
the dynamical systems describing DMMs are {\it integrable}, namely all global unstable manifolds in their phase space are well-defined topological 
manifolds~\cite{Gilmore}.

\begin{figure*}[t!]
	\includegraphics[width=18cm]{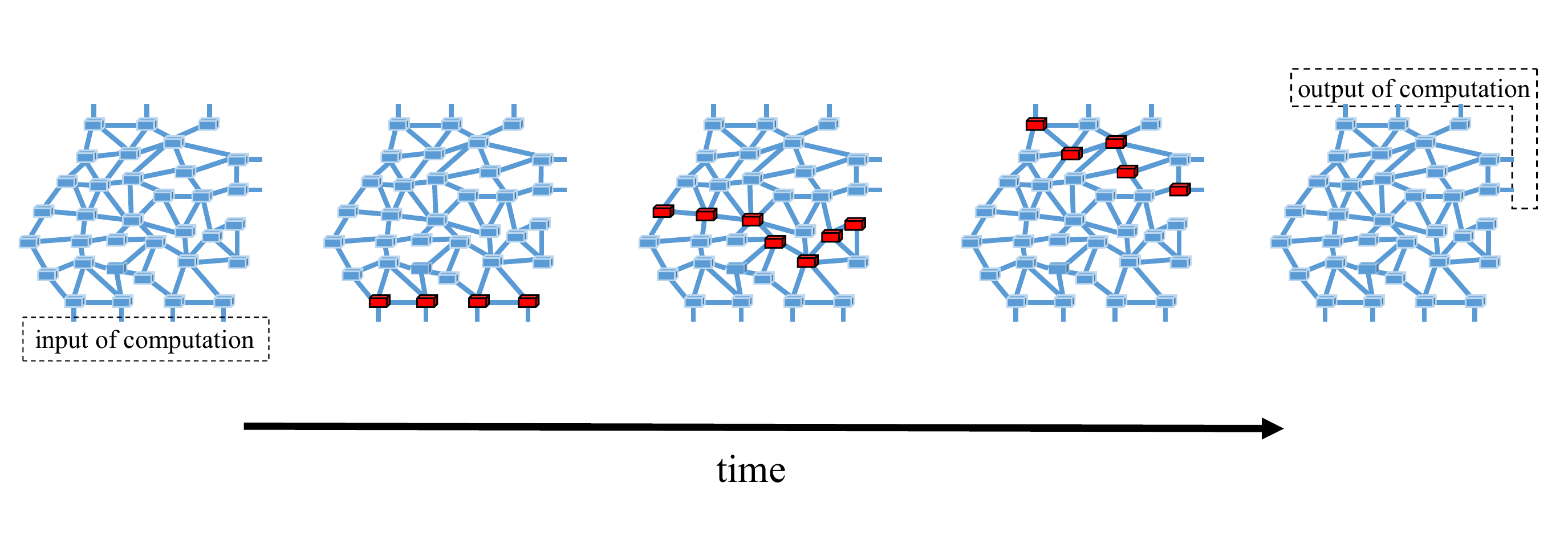}
	\caption{Schematic of the operation of a DMM as a physical system. The DMM is represented as a collection of elementary logic gates (blue rectangular 
		boxes) connected together to represent the original combinatorial optimization problem. At the initial time an input representing the problem instance is fed at the 
		appropriate terminals and the DMM has a number of ``logical defects'' (indicated by the red rectangular 
		boxes). The presence of logical defects indicates that the DMM logic circuit is not stationary, and thus it begins an instantonic
		dynamics towards one of the equilibrium states (local supersymmetric vacua). The solution of the original problem is then read at the appropriate 
		terminals at the end of the computation. The DMM dynamics can be viewed as an instantonic process during which the logical
		defects are being pushed out of the circuit.} 
	\label{figscheme}
\end{figure*}

Once the dynamical system that solves a specific problem has been identified, it can either be implemented in hardware or, since the corresponding equations of motion~(\ref{ODE}) represent a non-quantum system, they can be efficiently integrated numerically in time to find the equilibrium points. At equilibrium, the internal memory variables define center manifolds (``flat directions'' in the phase space) and effectively decouple from the voltage variables~\cite{DMM2}. The presence of center manifolds is simply a manifestation of the fact that once all SOLGs are logically satisfied (i.e., all voltages have reached the values representing either the logical 
1 or the logical 0), the internal memory variables loose their purpose, and the gates are logically-consistent irrespective of the values acquired by the internal variables. The approach to equilibrium of DMMs would then be enough to solve the original problem (see schematic operation of a DMM in Fig.~\ref{figscheme}).

In fact, it has already been shown that the {\it simulations} of the equations of motion of DMMs perform orders of magnitude faster than traditional algorithmic approaches on a wide variety of combinatorial optimization problems~\cite{exponential2017speedup,DMMperspective,AcceleratingDL,ILP,spinglass}. To better understand the physical reason behind this efficiency, in Ref.~\cite{topo} we have investigated the transient dynamics of DMMs by employing the newly developed (supersymmetric) topological field theory (TFT) of dynamical systems~\cite{Entropy}. In that work we have shown that the transient dynamics of DMMs proceed via 
elementary {\it instantons}, namely through classical trajectories that connect critical points in the phase space with different stability (indexes). The collective character of the instantons, expressed in their fermionic zero modes, is then at the origin of the {\it dynamical long-range order} (DLRO) in DMMs. 
It is this DLRO that allows these machines to compute complex problems efficiently.

In the present work, we aim at advancing further our understanding of the workings of DMMs and the close relation between logic, dynamics, and topology that the concept of DMMs conveniently offers. In particular, we discuss the ``dynamical'' meaning of the instantonic matrix elements as intersection numbers on instantons. We also argue that since the total number of elementary instantons cannot exceed the number of state variables of DMMs, which in turn can only grow at most polynomially with the problem size, the expected number of instantonic steps required to reach equilibrium can only grow at most polynomially. 

The structure of the paper consists of two major parts. First, we briefly discuss the supersymmetric theory of dynamical systems necessary to describe the dynamics of DMMs (Sec.~\ref{TFT}). Next, we discuss in Sec.~\ref{Instantons} the topological matrix elements on instantons and their relation to intersection numbers. We finally conclude in Sec.~\ref{Conclusions}. 

\section{Supersymmetric Topological Field Theory}\label{TFT}

In order to accomplish the above program, we realize that a deeper insight into the dynamics represented by Eqs.~(\ref{ODE}) can be obtained if, instead of working directly with the trajectories $x(t)$ in the (non-linear) phase space $X$, we transition to an {\it algebraic representation of dynamics}. If we do that, we can then mirror the 
algebraic structure of Quantum Mechanics and take advantage of vector algebra and calculus in topological vector spaces (Hilbert spaces). 

In other words, as in Quantum Mechanics, we want to study the dynamics~(\ref{ODE}) by means of {\it state vectors} $\ket{\psi}$ in, and operators on a linear {\it Hilbert space}. This Hilbert space is the (complex-valued) exterior algebra (or Grassmann algebra) of $X$: $\Omega(X,\mathbb{C})$~\cite{DiffGeometry}.

This algebraic program was pioneered by Koopman~\cite{Koopman} and von Neumann~\cite{vonNeumann1,vonNeumann2} in the 1930s for Hamiltonian systems. It 
was later discussed in the context of ``stochastic quantization''~\cite{ParSour}, revealing the supersymmetric structure of the theory~\cite{Gozzi0}. Finally, it was recently extended to generic dynamical systems, whether deterministic or stochastic, establishing that a supersymmetric TFT encompasses 
all classical (noisy) dynamical systems~\cite{Entropy}. The supersymmetric TFT that emerges is a type of cohomological field theory~\cite{book,Witten1,Witten2,labastida1989,Anselmi_1,TQM,Frankel} with the Noether charge of the {\it topological supersymmetry} (TS) being the exterior derivative, $\hat d$, on the exterior algebra of the phase space. 

If unbroken, this TS represents the preservation of the phase-space continuity of all dynamical systems. A spontaneously broken TS, instead, is the algebraic essence of phenomena such as chaos, 1/f noise, etc., with associated power-law correlations. In this section we report only the main ingredients of this 
theory needed to understand the results of Sec.~\ref{Instantons}. An extensive review of these concepts can be found in Ref.~\cite{Entropy}.

Let us start by noting that although DMMs are {\it deterministic} dynamical systems, any physical system is always subject to some external noise. It is then meaningful to add to 
Eqs.~(\ref{ODE}) some degree of noise. One can always take the limit of zero noise at the end of the procedure. (Adding the noise to the system has also the mathematical advantage of making the stochastic evolution operator (Eq.~(\ref{SEO_PathIntegral}) below) elliptic.) We therefore perturb the equations of motion of DMMs by a stochastic noise:
\begin{eqnarray}\label{stochODE}
F \to \tilde F = F + (2\Theta)^{1/2} \sum\nolimits_{a=1}^k e_a(x)\xi^a(t), 
\end{eqnarray}
where $\{e_a(x)\in TX_{x}, a=1...k\}$ is a collection of vector fields that are responsible for coupling of the noise to the system with $TX_x$ being the tangent space of $X$ at point $x$, $\Theta$ is the intensity of the noise, and $\xi^a(t)\in \mathbb{R}$ is a noise variable that may be assumed to be Gaussian white. We can now follow a type of stochastic quantization procedure~\cite{ParSour,ZinnJustin} which can be identified as a version of the  Becchi-Rouet-Stora-Tyutin (BRST) quantization applied to a stochastic differential equation (SDE). 

\subsection{Topological Viewpoint}

Let us start by discussing some of the topological aspects of this procedure and inspect the following object, 
\begin{eqnarray}
w(\xi) = \int_{p.b.c.} Dx\prod_\tau \delta(\dot x(\tau) - { \tilde F}(\tau))J,\label{Object}
\end{eqnarray}
where $\xi$ is a single realization of the noise, and the functional integration is performed over closed paths and/or periodic boundary conditions (p.b.c.), $x(t)=x(t')$. The functional determinant or the Jacobian is
\begin{eqnarray}
J = \operatorname{Det}\frac{\delta(\dot x(\tau) - { \tilde F}(x(\tau)) )}{\delta x(\tau')}.
\end{eqnarray}

The $\delta$-functional reduces the functional integration over closed paths to the summation over the closed solutions of the SDE only,
\begin{eqnarray}
w(\xi) = \sum_{\text{closed solutions}} \operatorname{sign} J.\label{Object_sum}
\end{eqnarray}
One would expect that these closed solutions as well as their total number are dependent on the noise configuration. In reality, $w(\xi)$ has a topological character.  

We can see this from the following analogy with finite-dimensional vector fields. Namely, we can think of $\dot x(\tau) - { \tilde F}(x(\tau))$ in Eq.~(\ref{Object_sum}) as a vector field on the space of all closed paths. Then, the right hand side of Eq.~(\ref{Object_sum}) is the summation over the indices (the signs of determinants of the Jacobians) of the critical points of this vector field, i.e, for those trajectories such that $\dot x(\tau) - { \tilde F}(x(\tau))=0$. In finite-dimensional (orientable compact) spaces, such sum would equal the Euler characteristic of the space according to the Poincar\'e-Hopf theorem (see, e.g., Ref.~\cite{Katok-1}). An infinite-dimensional generalization of the concept of Euler characteristic is not so straightforward from a mathematical point of view. Fortunately, we do not need such a generalization. All we need from the above analogy is understanding that $w$ is of topological nature and it does not depend on the configuration of the noise. 


Now, since $w(\xi)$ is independent of $\xi$, we can introduce the stochastic average of $w$,
\begin{eqnarray}
{W} = \left\langle w \right\rangle,\label{WittenIndex}
\end{eqnarray}
where the angled brackets denote stochastic averaging over the noise configurations,
\begin{eqnarray}
\langle A(\xi) \rangle = C^{-1}\int D \xi A(\xi) P(\xi), 
\end{eqnarray}
with $A(\xi)$ being some functional of $\xi$ and $P$ is the probability distribution of noise configurations with $C=\int D \xi P(\xi)$ being a normalization constant.

\subsection{Gauge-Fixing Perspective}
Substituting Eq.~(\ref{Object}) into Eq.~(\ref{WittenIndex}) and using standard path-integral techniques~\cite{Book_Peskin} that allow to exponentiate 
the bosonic $\delta$-functional and the functional determinant, one obtains the following expression
\begin{eqnarray}
W = \left\langle \int_{p.b.c.} D\Phi e^{\{ Q, i\int_{t'}^t d\tau \bar \chi(\tau)(\dot x(\tau)-\tilde F(x(\tau))) \}} \right\rangle,\label{Witten_Path_1}
\end{eqnarray}
where the notation $\Phi$ is introduced for the collection of the original bosonic fields, $x$, and additional fields that are the bosonic momentum (also known as Lagrange multiplier), $B_i$, and a pair of Faddeev-Popov ghosts, $\chi^i$ and $\bar\chi_i$. The operator of BRST symmetry that can be defined via its action on an arbitrary functional $A$ is (summation over repeated indexes is understood)
\begin{eqnarray}
\{  Q, A\} = \int_{t'}^t d\tau \left(\chi^i(\tau)\frac{\delta }{\delta x^i(\tau)} + B_i(\tau)\frac{\delta }{\delta \bar\chi_i(\tau)}\right) A. \label{Q_pathint}
\end{eqnarray}
Note also that all the fields in the above path-integral have periodic boundary conditions including fermions for which periodic boundary conditions are unphysical. This is the path-integral origin of the alternating sign operator in the operator expression of the Witten index~\cite{Witten1,Witten2}.

The functional integration over Gaussian white noise can now be performed exactly and this leads to 
\begin{eqnarray}
W = \int_{p.b.c.} D\Phi \; e^{ \{Q,\Psi\} },\label{Witten_Path_2}
\end{eqnarray}
with the so-called gauge-fermion 
\begin{eqnarray}
\Psi = \int^{t}_{t'} d\tau\left(i \bar\chi_j(\tau) \dot x^j(\tau) - \bar d(\Phi(\tau))\right),
\end{eqnarray} 
and 
\begin{eqnarray} 
\bar d(\Phi) = i\bar\chi_j\left(F^j - \Theta e_a^j \{{  Q}, i\bar \chi_ke_a^k\}\right), \label{Current_Op_Path}
\end{eqnarray}
that can be loosely identified as the ``probability current''. 

It is no accident that integrating out the noise leaves the action of the model $Q$-exact, i.e., of the form of $\{Q , \cdot \}$. This is due to the nilpotency of the BRST operator, $\{ Q, \{Q , \cdot \} \}=0$, which is the path-integral version of the nilpotentcy of the exterior derivative on the 
exterior algebra of the phase space (see below). It is the nilpotency of $Q$ that renders the cross-term, $\{Q, i\bar\chi_je_a^j \}\{Q, i\bar \chi_ke_a^k\}$, a $Q$-exact piece $\{ Q,i\bar\chi_je_a^j \{Q, i\bar \chi_ke_a^k\} \}$, which, in turn, assures that the whole action is $Q$-exact.

In high-energy models, the functionality of $Q$-exact terms is to fix gauges~\cite{Book_Peskin}. The BRST symmetry generates fermionic versions of gauge transformations and the overall effect of the BRST-exact terms in the action is to limit the path-integration to only paths that satisfy the gauge condition. The same interpretation applies to our case here. The BRST symmetry generates (the fermionic version of) all possible deformations of the path, $\{ Q, x(\tau) \}=\chi(\tau)$, and the $Q$-exact action leaves out only solutions of the SDE.

\subsection{Topological Field Theory Viewpoint}



The stochastic evolution operator (SEO) of the theory can be obtained simply by removing the periodic boundary conditions in the path-integral representation of the Witten 
index $W$:
\begin{eqnarray}
\hat {M}_{tt'}(x\chi,x'\chi') = \iint_{\genfrac{}{}{0pt}{}{x(t)=x,x(t')=x'}{\chi(t)=\chi,\chi(t')=\chi'}} D\Phi \; e^{\{{\mathcal Q},\Psi\}}\label{SEO_PathIntegral},
\end{eqnarray}
where the path integration in Eq.~(\ref{SEO_PathIntegral}) is over trajectories connecting the arguments of the evolution operator.



The operator representation of the SEO is
\begin{eqnarray}
\hat {M}_{tt'}(x\chi,x'\chi') = e^{-(t-t')\hat H} \delta(x-x')\delta(\chi-\chi')\label{SEO_Oper}.
\end{eqnarray}
Here, the infinitesimal SEO, $\hat H$, is  
\begin{eqnarray}
\hat H=\hat L_{F} + \Theta \sum\nolimits_{a=1}^k\hat L_{e_a}\hat L_{e_a},
\end{eqnarray}
where the operators $L_i$ are Lie derivatives over the vector fields indicated by their subscripts. The meaning of $\hat H$ is simple. The first term is the flow along $F$ while the second term is the noise-induced diffusion.

The infinitesimal SEO can be given also the explicitly supersymmetric form
\begin{eqnarray}\label{HEO}
\hat H=[\hat d, \hat{\bar d}],
\end{eqnarray}
where $\hat d$ is the exterior derivative, and the current $\hat{\bar d} = \imath_{F} + \Theta \sum\nolimits_{a=1}^k\imath_{e_a}\hat L_{e_a}$ is the operator version of Eq.~(\ref{Current_Op_Path}), with $\hat\imath_i$  interior multiplications. The 
square bracket $[  \cdot , \cdot ]$ denotes the bi-graded commutator of any two arbitrary operators. The exterior derivative is the operator version of the topological sypersymmetry $Q$ in the path integral~(\ref{Witten_Path_2}). 

The exterior algebra of the phase space $X$
\begin{equation}
\Omega(X,\mathbb{C})=\bigoplus_{k=0}^D \Omega^k(X,\mathbb{C}), 
\end{equation}
is the vector space of differential $k$-forms of all degrees 
\begin{equation}\label{differ}
\ket{\psi^{(k)}(x)}=\frac{1}{k!}\sum_{i_1\dots i_k}\psi_{i_1\dots i_k}^{(k)}(x) dx^{i_1}\wedge \dots \wedge dx^{i_k} \in \Omega^{(k)}(X,\mathbb{C}),
\end{equation}
where $\psi_{i_1\dots i_k}^{(k)}(x)$ is a smooth antisymmetric tensor, $\wedge$ is the wedge product (or multiplication) of differentials $dx^{i}$, e.g., $dx^1\wedge dx^2 = dx^1 \otimes dx^2 - dx^2 \otimes dx^1$. The space $\Omega^k(X,\mathbb{C})$ is that of 
all differential forms of degree $k$. In this language the exterior derivative is simply:
\begin{equation}\label{dext}
\hat d = \sum_{i=0}^D dx^i \wedge \frac{\partial}{\partial x^i}.
\end{equation}

Finally, any state of the system in the exterior algebra language can be represented as  
\begin{equation}\label{state}
\ket{\psi(x)}=\sum_{k=0}^D \ket{\psi^k(x)},
\end{equation}
which concludes our original program of studying the dynamics~(\ref{ODE}) by means of state vectors $\ket{\psi}$ in, and operators on a linear Hilbert space.

Let us note, however, an important difference with Quantum Mechanics: the evolution operator, $\hat H$, is {\it pseudo-}Hermitian because all its entries are real~\cite{Mos023}. This means that its spectrum is composed of only real eigenvalues and pairs of complex conjugate eigenvalues. The closest analogue of complex conjugate pairs from the dynamical systems theory would be the Ruelle-Pollicott resonances~\cite{E_Ruelle}. 

In addition, from Eq.~(\ref{HEO}), we see that the evolution operator, $\hat H$, is $\hat d$-{\it exact}. Since the exterior derivative $\hat d$ is nilpotent, $\hat d^2 =0$, it commutes 
with any $\hat d$-exact operator: $[\hat d, [ \hat d, \hat X]]=0, \forall \hat X$. This means that the evolution 
operator commutes with $\hat d$:
\begin{equation}
[\hat H, \hat d] =0,
\end{equation}
and $\hat d$ is therefore a symmetry of the system. 

From Eq.~(\ref{dext}), and the fact that the wedge product $\wedge $ is antisymmetric, we also see that $\hat d$ creates {\it fermionic} or anti-commuting variables ($dx^i \wedge \equiv \chi^i$), and destroys {\it bosonic} or commuting variables ($x^i$). Therefore, $\hat d$ is a {\it supersymmetry} of the dynamical system. On the other hand, $\hat {\bar d}$ is not necessarily nilpotent and hence, apart from specific model systems~\cite{ParSour}, it does not generally commute with $\hat H$.

\section{Topological Features of DMMs}\label{Instantons}

We have therefore shown that Eqs.~(\ref{stochODE}), as well as their deterministic limit, Eqs.~(\ref{ODE}), are members of the family of (Witten-type) 
cohomological field theories. This means that certain objects are topological invariants.

The quantity~(\ref{Witten_Path_1}) is one of such objects. It is the famous Witten index~\cite{Witten1,Witten2} that in our case and in operator representation has the following form:
\begin{eqnarray}
W=\operatorname{Tr } (-1)^{\hat F}e^{-(t-t')\hat H}.
\end{eqnarray}
where $\hat F$ is the ghost/fermion number operator.

Another interesting class of topological invariants is formed by the matrix elements on instantons. From a technical point of view, an instanton is a family of classical solutions of a deterministic equation of motion that starts at an unstable critical point, or a saddle point, and ends at a more stable critical point~\cite{Coleman}
\begin{eqnarray}
\dot x_{cl}(t,\sigma) = F(x_{cl}(t,\sigma));\;\;\;\;\;\;  x_{cl}(\pm\infty,\sigma)=x_{i,f},\label{instanton}
\end{eqnarray}
with $x_i$ and $x_f$ the initial and final critical points, respectively, of the flow vector field $F$ with different indexes. The parameters $\sigma$ are the so-called modulii of instantons, and represent the {\it non-local} (collective) character of instantons~\cite{Coleman}.

Such objects are manifolds (with boundary) of dimensionality equaling the difference in indixes of the initial and final critical points~\footnote{The index of a critical point is the number of its unstable dimensions.}. Indeed, Witten-type topological field theories are sometimes identified as intersection theory on instantons understood as these manifolds~\cite{Book1}.

From a physical point of view, processes such as earthquakes, solar flares, avalanches, etc. can be identified as instantons. Another example of instantonic dynamics is quenches. After some external parameters of a dynamical system are suddenly changed, the latter finds itself in an unstable point, and begins its dynamics to a new stable point. It is then not so surprising that the operation of a DMM is an example of post-quench instantonic dynamics where the dynamics-initiating quench is the assignment of new input variables representing the logic problem that needs to be solved (see Fig.~\ref{figscheme}).

\subsection{Instanton matrix elements}

To show explicitly that some instanton matrix elements are topological invariants  we then compute correlators on instantons for appropriate observables. For a set of $l$ observables $O_{\alpha_j}(\hat \Phi)$ 
(with $\hat \Phi$ the operators of the corresponding fields) we can then compute matrix 
elements of the type
\begin{eqnarray}\label{Instantonop}
{\mathcal I} = \langle f| {\cal T}\prod\nolimits_{j=1}^l O_{\alpha_j}(\hat \Phi(t_j))|i\rangle,
\end{eqnarray}
where the operators are in the Heisenberg representation, $\hat \Phi (t)=\hat {M}_{0t}\hat \Phi \hat {M}_{t0}$, with $\hat \Phi$ being Schroedinger operators. Since the choice of the reference time instant is irrelevant we have taken it to be zero, and $\cal T$ denotes the operator of chronological ordering. 
\begin{figure}[t!]
	\includegraphics[width=1.02\columnwidth]{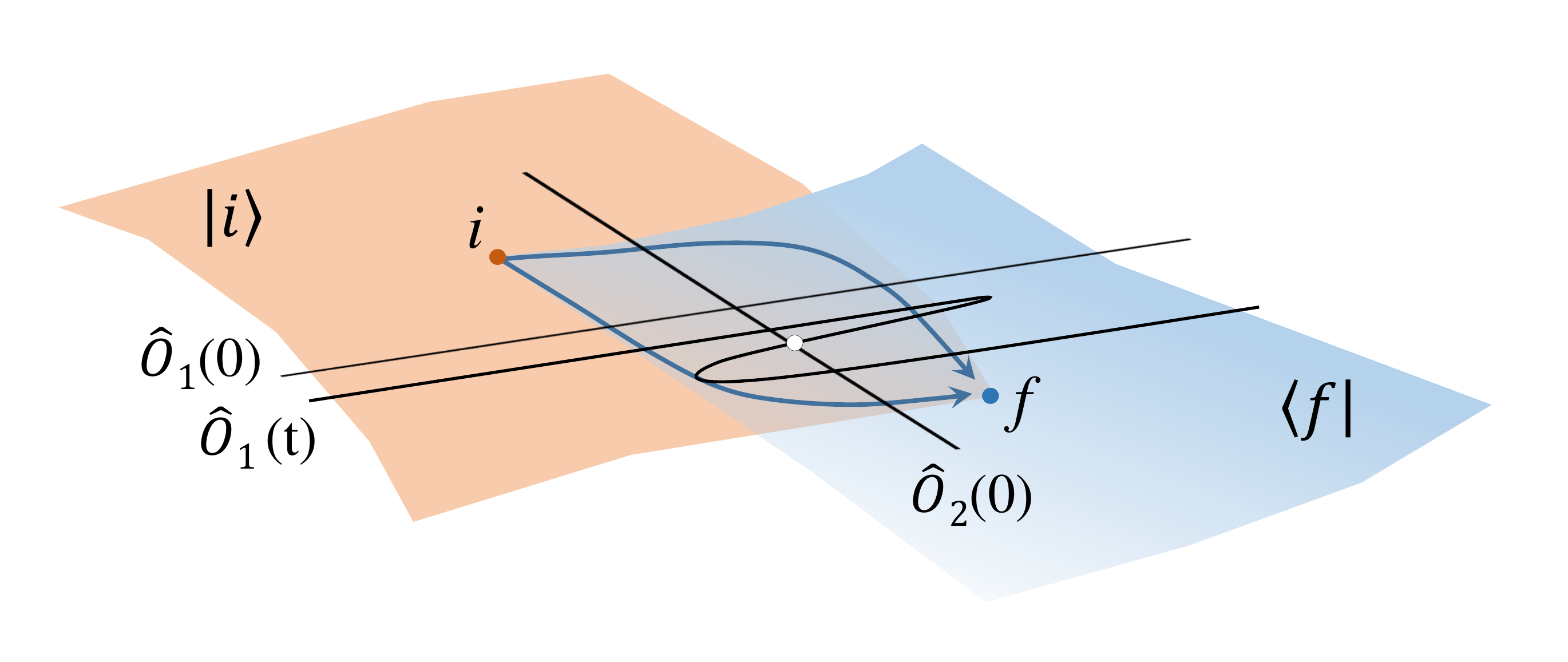}
	\caption{Schematic representation of the instantonic matrix element, Eq.~(\ref{finalI}), with only two BPS observables of the type $O_{\alpha}(\hat \Phi)=\delta(\hat x^{\alpha}-1/2)\hat \chi^{\alpha}$. The bra of the $f$-vacuum, $\langle f|$, is the Poincar\'e dual of the attraction basin of the critical point $f$ (light blue area), and the ket of the $i$-vacuum, $|i\rangle$, is the Poincar\'e dual of the repulsion basin of the critical point $i$ (orange area). The instanton, $\langle f|...|i\rangle$, is the overlap that consists of all trajectories (curved blue arrows) starting at $i$ and ending at $f$. The ``on-time'' instantonic matrix element $\langle f|\hat O_1(0)\hat O_2(0)|i\rangle=1$, is the intersection number of the two hyperplanes, $x^{1,2}=1/2$. Even if the time argument of $\hat O_1(t)$ changes, the intersection number remains the same since new intersections appear in pairs of positive and negative (white circle) Jacobians, thus canceling each other in Eq.~(\ref{finalI}).} 
	\label{figint}
\end{figure}

The states $\langle f |$ and $| i \rangle $ are the bra and ket of the $i$- and $f$-vacua, i.e., supersymmetric perturbative ground states, associated with the respective critical points, $x_i$ and $x_f$. They are the Poincar\'e duals of the local stable and unstable manifolds of the critical point for the bra and ket of the vacuum, respectively (see Fig.~\ref{figint}), namely differential forms that are constant functions (without fermions) along the
manifolds, and $\delta$-function distributions with fermions 
in the transverse directions. In view of their different fermionic content, their overlap is zero as can be easily determined from the path integral:
\begin{eqnarray}\label{overlap}
\int_{x(-\infty)=x_{i}, x(+\infty)=x_{f}, }D\Phi e^{S(\Phi)} = \langle f| i \rangle = 0.
\end{eqnarray}

The matrix elements~(\ref{Instantonop}) are topological invariants only for Bogomol'nyi-Prasad-Sommerfield (BPS) observables, namely those defining the topological sector. Non-BPS observables would contribute short-ranged correlators, and therefore would not reveal 
the topological character of the theory in the long-time limit. As BPS observables relevant to DMMs we can choose the operators $O_{\alpha}(\hat \Phi)=\delta(\hat x^{\alpha}-1/2)\hat \chi^{\alpha}$~\cite{topo}. These observables can be interpreted as ``detecting" when the voltages 
on the terminals of the gates ``cross" the value 1/2, either towards the logical 0 or the logical 1. The missing ghost (fermion) in each  
unstable direction of the initial state is compensated by a fermion $\chi^{\alpha}$. 

It is worth noting that the presence of ghosts in the observables should not be surprising: for consistency, the observables themselves need to be represented using the same supersymmetric TFT used to describe the system of interest. Without the correct number of fermions (which must be equal to the dimension $d$ of the modulii space) the matrix element~(\ref{Instantonop}) would be 
identically zero (see, e.g., Eq.~(\ref{overlap})). 

The calculation of $\mathcal I$ is then done in the standard manner~\cite{Book1}. First, we take advantage of the fact that the supersymmetric description is coordinate-free. We can then choose as coordinates the instantonic modulii, and fluctuations around them,
\begin{eqnarray}
x^i(t) &=& x^i_{cl}(t,\sigma) + ...,
\end{eqnarray}
where the dots represent all the other fluctuational modes. Each modulus provides one {\it fermionic zero mode} to the deterministic equations of motion for the fermions,
\begin{eqnarray}\label{Instantderiv}
(\hat \partial_t - TF) \lambda_j(t,\sigma) = 0;\;\;\;\; \lambda^i_j(t,\sigma) = \frac{\partial x_{cl}^i(t,\sigma)}{\partial \sigma^j },
\end{eqnarray}
where $TF^i_k=\partial F^i/\partial x^k$. Eq.~(\ref{Instantderiv}) can be obtained by differentiating Eq.~(\ref{instanton}) over $\sigma^j$ once. 

Zero modes must be given a special care. We then introduce the supersymmetric partners of the modulii, $\nu^{j}$, 
\begin{eqnarray}
\chi^i(t) &=& \lambda^i_j(t,\sigma) \nu^{j} + ...,
\end{eqnarray}
where once again the dots represent all the other modes. Then, one has 
\begin{eqnarray}\label{instantmod}
{\mathcal I} &=& \int \prod_{i=1}^{l} d\sigma^i d\nu^{i} \delta(x^{\alpha_{i}}_{cl}(t_i,\sigma)-1/2)\nonumber\\
&&\times \left(\lambda^{\alpha_i}_j(t_i,\sigma) \nu^{j} + ...\right)\iint D\Phi'e^{\{{\mathcal Q}, \Psi^{(2)}(\Phi')\}},
\end{eqnarray}
The path-integral in the second line of Eq.~(\ref{instantmod}) is over all the other modes, and only the Gaussian part of the action (one loop) is left in the exponent (the term $\{{\mathcal Q}, \Psi^{(2)}(\Phi')\}$). Such integrals are always unity due to the supersymmetric cancellation of the fermionic and bosonic determinants (``localization principle'' of supersymmetric theories)~\cite{Book1}. This is the reason why the one-loop approximation is exact in the present case. 

Therefore, one is left with
\begin{eqnarray}\label{finalI}
{\mathcal I} = \sum_{\sigma_0, x_{cl}^{\alpha_i}(t_i,\sigma_0)=1/2} \text{sign Det} \left.\frac{\partial x_{cl}^{\alpha_i}(t_i,\sigma)}{\partial \sigma^j}\right|_{\sigma=\sigma_0},
\end{eqnarray}
which is a {\it topological invariant}. 

\subsection{Intersection theory on instantons}

We can now see where the DLRO originates from. Since the BPS observables we have chosen are the Poincar\'e duals of the hyperplane $x^{\alpha}=1/2$, the matrix element $\mathcal I$ can be interpreted as an {\it intersection} of a collection of such hyperplanes on the instanton. In fact, $\sigma_0$ is the point of the modulii space where all variables $x_{cl}^{\alpha_i}$ acquire the value 1/2. In turn, $\mathcal I$ is invariant no matter how far the terminal voltages in the DMM are separated from each other spatially. This DLRO then originates from the spatial nonlocality of the collective instantonic variables (the instanton modulii). 

Furthermore, $\mathcal I$ is also independent of time variables: if the time argument of the observables in Eq.~(\ref{Instantonop}) changes, pairs of solutions with positive and negative Jacobians appear, canceling each other in Eq.~(\ref{finalI}). This is explicitly shown in Fig.~\ref{figint} for 
two observables. In fact, the ``on-time'' instantonic matrix element 
\begin{equation}\label{Intersect}
\langle f|\hat O_1(0)\hat O_2(0)|i\rangle=1,
\end{equation}
is the {\it intersection number} of the two hyperplanes, $x^{1,2}=1/2$ and cannot change even if the time argument of, say, $\hat O_1(t)$ changes. Again, this 
is because new intersections appear in pairs of positive and negative Jacobians, thus canceling each other in Eq.~(\ref{finalI}).

This demonstrates that the transient dynamics of DMMs has also a temporal long-range order. This temporal order is due to the fact that the initial state of the dynamics has unstable variables, and thus the trajectory is highly sensitive to initial conditions. On the other hand, the solution search is robust against perturbations and initial conditions~\cite{Bearden}, due to the topological character of the critical points whose index and number cannot change by perturbative effects.

\subsection{DLRO vs. chaos}

It is also worth stressing that the DLRO due to the {\it collective} instantonic variables is {\it not} the same as the one due to the spontaneous breakdown 
of TS. In the former case, the low-energy part of the liberated dynamics is of an ``avalanche'' type connecting different {\it perturbative} vacua. On instantons, the TS  
is {\it effectively} (although not globally) broken giving rise to DLRO. This order can be interpreted as due to the release of {\it goldstinos} (fermionic Goldstone modes) every time the system transitions from one local vacuum to another. 

On the other hand, in the case of {\it spontaneous} breakdown of supersymmetry, $\bra{\textrm{gs}}\hat H\ket{\textrm{gs}}\neq 0$, namely the exterior derivative does not annihilate the {\it global} vacuum, $\ket{\textrm{gs}}$: $\hat{d}\ket{\textrm{gs}}\neq 0$. The corresponding liberated dynamics would then consist of a sea of gapless goldstinos that are unable to restore the supersymmetry: the system is unable to thermalize, and, therefore, it must show chaotic behavior. 

As already mentioned, DMMs never break (global) supersymmetry, hence do not support chaotic dynamics: the corresponding dynamical system is integrable~\cite{topo}. Since integrability means that all global unstable manifolds (GUMs) in the phase space are well-defined topological manifolds~\cite{Gilmore}, the Poincar\'e duals of GUMs are the eigenstates with zero eigenvalue and the operator $\hat d$ annihilates them (they are $\hat d$-closed)~\cite{Igor}. Since $\hat d$ is the operator version (in cohomology) of the boundary operator (in homology), this means that GUMs in DMMs have no boundaries. 

\subsection{Solitons and logical defects}
Instantons are solitons in one lower dimension~\cite{Solitons}, therefore, each instanton can be interpreted as corresponding to the elimination of {\it solitonic configurations} of logical inconsistency (``logical defects'') from the circuit (see schematic in Fig.~\ref{figscheme}). Since the initial state of the dynamics has a certain number of logical defects, during the transient phase a DMM attempts to rid itself of these defects till the solution is found. 

In order to draw a parallel between DMMs and other physical models, the following analogy with the 1D Ising ferromagnet may be useful. A finite 1D chain of atoms, each having spins either up or down, has two ground states or vacua: one corresponding to all spins up, the other to all spins down. In order to switch the system from one vacuum (say, all spins up) to the other (all spins down), one can force the rightmost atom in the chain to flip its spin. This operation is the direct analogue of providing the DMM with new input 
variables (cf. Fig.~\ref{figscheme}). In the case of the 1D spin chain, the operation of flipping the rightmost spin then creates a soliton called a kink or domain wall. What happens next is the instantonic process of ``killing'' this soliton: the soliton travels through the system from one side to the other and exits the chain, leaving the system at the other vacuum. Similar considerations are valid also in higher dimensions. The important point in the above analogy is the necessity to create a soliton that the system has to instantonically ``push out'' of itself to switch to a different vacuum. 

It is also important to note that, strictly speaking, solitons are defined in continuous space models or on lattices that allow a ``coarse-graining'' procedure and with interactions between nearest neighbors only~\cite{Solitons}. It is for this reason that solitons typically have finite dimensions. For example, the domain wall that separates regions of different vacua must have a finite width. DMMs, on the other hand, represent logical circuits. Therefore, they are almost never structured lattices that favor a ``coarse-graining'' procedure. It is for this reason that, properly speaking, logical defects in DMMs are in fact generalizations of the classical concept of solitons. In particular, it is possible that in certain situations a solitonic configuration (a logical defect) may occupy the entire circuit~\cite{spinglass}. 

\subsection{Instantons and steps to solution}

Finally, we can use the above arguments to count the total number of instantonic steps, $\cal N$ (the dimensionality of the composite instanton), a DMM requires to reach the solution of a given 
problem. We first recall from the Introduction that DMMs, if properly designed, are point-dissipative systems~\cite{hale_2010_asymptotic}, namely all trajectories of the system are {\it bounded} and will eventually end up into one of the attractors, irrespective of the initial conditions. 

Now, instantons can only connect critical points of a given stability with critical points that are more stable~\cite{Coleman}. Since the number of unstable directions is 
at most equal to $D$ (the dimensionality of the phase space), and the latter can only grow polynomially with problem size~\cite{DMM2,DMMperspective}, the 
total number of instantonic steps to reach equilibrium can only grow polynomially with system size: 
\begin{equation}
{\cal N} \sim {\mathcal O}({\cal P}(D)).
\end{equation}
As previously 
mentioned, these instantonic steps correspond to the elimination of solitonic configurations of logical defects. Therefore, a DMM is able to reach 
solution by eliminating a set of logical defects that can only grow polynomially with problem size. 


\section{Conclusions}\label{Conclusions}
In conclusion, we have provided further theoretical understanding of the operation of digital memcomputing machines~\cite{DMM2,DMMperspective}: a novel class of computinational machines specifically 
designed to tackle combinatorial optimization problems. The physical (electrical circuit) realization of these machines gives rise to a set of non-linear differential equations for the voltage and internal (memory) variables. These equations, in turn, can be described algebraically using a (supersymmetric) topological field theory~\cite{Entropy}. 

This TFT has revealed that the transient dynamics of these machines is a composite instanton connecting critical points of different indexes in the phase space. The topological supersymmetry is effectively broken on instantons, although it is never globally broken, implying absence of chaotic behavior: DMMs are integrable systems. A DMM then finds the solution of the original problem via a succession of elementary instantons whose role is to eliminate solitonic configurations of logical defects from the circuit.

The {\it collective} character of the instantons connecting these critical points is responsible for the DLRO of DMMs, as we have explicitly shown by computing correlators on instantons within the topological sector of the theory. We have also argued that the dimensionality of the composite instanton cannot exceed the number of state variables of DMMs, which in turn can only grow at most polynomially with the size of the problem. 

These studies further highlight the topological ({\it collective}) dynamical behavior of DMMs. These properties turn out to be key for their ability to 
solve hard problems efficiently. This work  
also reinforces the notion that physics-based approaches to computation offer advantages that are not easily obtained via traditional algorithmic means. 

\bibliography{SUSYref}

\end{document}